\documentclass[useAMS,usenatbib,fleqn]{mn2e}
\usepackage{graphicx,epsfig}
\usepackage{amsmath}
\usepackage{booktabs}
\usepackage{multirow}
\usepackage{multicol}
\usepackage{tabularx}
\usepackage{mathrsfs}

\def\gsim{\;\lower4pt\hbox{${\buildrel\displaystyle >\over\sim}$}\;}
\def\lsim{\;\lower4pt\hbox{${\buildrel\displaystyle <\over\sim}$}\;}
\def\grls{\;\lower4pt\hbox{${\buildrel\displaystyle >\over <}$}\;}
\def\vec#1{\mathbf{#1}}                
\def\vd#1{\sigma^2_{#1}}               
\def\e#1{\mathrm{e}^{#1}}              
\def\fp#1{\langle F'_{#1} \rangle}     
\def\fpp#1{\langle F''_{#1} \rangle}   
\def\d{\mathrm{d}}                     
\def\Mpch{h^{-1}\,\mathrm{Mpc}}        

\title[Analytic model for redshift-space distortions]{An analytic
  model for redshift-space distortions}
\author[Wang et al.]{
  Lile Wang$^{1,2}$\thanks{Email: wll9004@gmail.com},
  Beth Reid$^{1}$, and
  Martin White$^{1,3}$ \\
  $^{1}$ Lawrence Berkeley National Laboratory, 1 Cyclotron Road,
    Berkeley, CA 94720, USA\\
  $^{2}$ Department of Physics and Tsinghua Centre for Astrophysics
 (THCA), Tsinghua University, Beijing 100084, China \\
  $^{3}$ Departments of Physics and Astronomy, University of
  California, Berkeley, CA 94720, USA
}
\date{Accepted 2012...; Received 2012...; in original form 2012}
\pubyear{2012}
\begin{document}

\maketitle
\begin{abstract}
  Understanding the formation and evolution of large-scale structure
  is a central problem in cosmology and enables precise tests of
  General Relativity on cosmological scales and constraints on dark
  energy.  An essential ingredient is an accurate description of the
  pairwise velocities of biased tracers of the matter field.  In this
  paper we compute the first and second moments of the pairwise
  velocity distribution by extending the Convolution Lagrangian
  Perturbation theory (CLPT) formalism of \citet{2012arXiv1209.0780C}.
  Our predictions outperform standard perturbation theory calculations
  in many cases when compared to statistics measured in N-body
  simulations.  We combine the CLPT predictions of real-space
  clustering and velocity statistics in the Gaussian streaming model
  of \citet{2011MNRAS.417.1913R} to obtain predictions for the
  monopole and quadrupole correlation functions
  accurate to 2 and 4 per cent respectively down to $<25\Mpch$
  for halos hosting the massive galaxies observed by SDSS-III BOSS.
  We also discuss contours of the 2D correlation function and
  clustering ``wedges''.  We generalize the scheme to
  cross-correlation functions.
  \end{abstract}
  
\section{Introduction}
\label{sec:introduction}

The large-scale structure (LSS) of the Universe, as traced for example
by the distribution of galaxies, is the focus of several ongoing and
upcoming observational campaigns.  In addition to furthering our
understanding of the cosmic web these projects seek to investigate
fundamental physics, including the properties of the initial
conditions, the imprint of (massive) neutrinos becoming
non-relativistic, and the behavior of the mysterious dark energy.
Improving our theoretical understanding of the LSS will enhance the
scientific return of these projects.  In particular, a more detailed
understanding of the anisotropy in the observed clustering is of great
interest, as the imprint of peculiar velocities in redshift survey
maps (known as redshift-space distortions: RSD) allows a consistency
test in general relativity between the expansion history and growth of
perturbations; such tests could provide support for modified gravity
theories as an explanation for the observed cosmic expansion.
Moreover, a precise understanding of the peculiar velocity induced
anisotropy in galaxy clustering would improve our ability to measure
the geometrically induced anisotropy known as the Alcock-Paczynski
effect \citep{1979Natur.281..358A}, and thus constrain the expansion
rate $H(z)$ directly \citep[for further details, see
e.g.][]{2011MNRAS.410.1993S}.

In this work we shall investigate an analytic model to predict the
two-point function of biased tracers of large-scale structure based on
perturbation theory.  There is a large literature using perturbative
techniques to study RSD \citep[see for example reviews
in][]{1998ASSL..231..185H, 2002PhR...367....1B,2009PhRvD..80d3531C}.
Standard perturbation theory (SPT) adopts an Eulerian description of
fluids, focusing on the velocity field and density contrast
(e.g.~\citealt{1980lssu.book.....P} for linear theory and
\citealt{1981MNRAS.197..931J,1983MNRAS.203..345V,1986ApJ...311....6G,
1992PhRvD..46..585M,1994ApJ...431..495J} for higher orders).  On the
other hand, Lagrangian perturbation theory performs an expansion in
the Lagrangian displacement field \citep{1992MNRAS.254..729B,
1994MNRAS.267..811B, 1995A&A...296..575B}. Lagrangian perturbation
theory (LPT) and SPT give identical results for the matter power
spectrum in real space when expanded to the same order
\citep{2008PhRvD..77f3530M}.  However, it is easier to include
redshift space distortions in LPT: a time derivative of the original
displacement field is simply added in the line-of-sight direction.
Furthermore, current theories of galaxy formation rely on the cooling
of gas within dark matter potential wells to form galaxies.
Therefore, like dark matter halos, galaxies are biased tracers of the
underlying matter distribution.  A local Lagrangian bias model seems
to provide a better description of dark matter halo clustering than a
local Eulerian bias
\citep[e.g.][]{2011MNRAS.415..829R,2012arXiv1201.4827B,
2012PhRvD..85h3509C, 2012PhRvD..86d3508W}, although additional terms
involving the tidal tensor may also become important for high mass
halos \citep{2012arXiv1207.7117S}.

Recently Lagrangian perturbation theory was extended by a resummation
scheme known as ``integrated perturbation theory''
\citep[iPT;][]{2008PhRvD..77f3530M,2008PhRvD..78h3519M}.  A key
success of iPT is a very accurate description of the redshift-space
two-point correlation function of dark matter halos on scales of
interest for studying baryon acoustic oscillations (BAO).
Unfortunately, the iPT predictions are inaccurate on scales
$20-70\Mpch$, even though deviations from linear theory are still only
$\sim 10$ per cent.

A recent paper by \citet{2012arXiv1209.0780C} introduced convolution
Lagrangian perturbation theory (CLPT) which improves the iPT method by
resumming more terms in the perturbative expansion.  CLPT gives
dramatically better results on small scales when compared to N-body
simulations, particularly for the description of the redshift-space
clustering of dark matter.  The methodology is easily extendable to
compute properties of the pairwise halo velocity distributions that
generate redshift-space distortions.  The primary purpose of this
paper is to examine CLPT's accuracy in predicting these statistics, in
comparison with N-body simulations.  We will see that CLPT provides an
accurate description of the velocity distributions.  Unfortunately,
the CLPT predictions for the anisotropy in the two-point correlation
function measured by the quadrupole are still inaccurate on the
quasi-linear scales of interest \citep[see figure 5
of][]{2012arXiv1209.0780C}.  Therefore, in this paper we combine the
real-space correlation function and the velocity statistics predicted
by CLPT with the non-perturbative approach advocated in \citet[][the
scale-dependent Gaussian streaming model]{2011MNRAS.417.1913R}.  This
model convolves the real-space two-point correlation function with an
approximation to the (scale-dependent) velocity distribution functions
to predict redshift space clustering.

This paper is structured as follows. In Sections \ref{sec:review} and
\ref{sec:v-sigma} we provide some analytic prerequisites for
evaluating clustering statistics with CLPT.  Section
\ref{sec:evaluate} contains the primary new calculation in this work
-- the prediction of pairwise velocity statistics for biased tracers
in CLPT.  We evaluate both auto- and cross-correlation statistics.  In
Section \ref{sec:GSM} we review the Gaussian streaming model, the
basis of our final model for the redshift space halo correlation
function.  In Section \ref{sec:results-statistics} we show the
numerical evaluation of those perturbative analytic results. Those
CLPT results are input into the Gaussian streaming model in Section
\ref{sec:redshift-space-stat} and compared with values obtained
directly and indirectly by simulations. Section \ref{sec:summary}
gives the summary of this article.

\section{Review}
\label{sec:review} Before we present our calculation of the velocity
moments in CLPT, let us review some background material to set our
notation and conventions.

\subsection{Background}

Throughout this work we will adopt the ``plane-parallel''
approximation, so that the line-of-sight (LOS) is chosen along a
single Cartesian axis ($\hat{\vec{z}}$).  While wide-angle effects
could potentially be important in modern surveys
\citep{2008MNRAS.389..292P}, \citet{2012MNRAS.420.2102S} have shown
that in practice these effects are small given current errors.  The
redshift-space position $\vec{s}$ of an object differs from its
real-space position $\vec{r}$ due to its peculiar velocity,
\begin{equation}
  \label{eq:sx}
  \vec{s} = \vec{x} + v_z(\vec{x})\,\hat{\vec{z}},
\end{equation}
where $v_z(\vec{x}) \equiv u_z(\vec{x})/(aH)$ is the
LOS component of object's velocity (assumed non-relativistic) in units
of the Hubble velocity.  In linear theory, the peculiar velocity field
is assumed curl-free, and its divergence is sourced by the underlying
matter fluctuations:
\begin{equation}
  \nabla \cdot \vec{v} = -f \delta_{m}
\end{equation}
where $f \equiv \d\ln D/\d\ln a$ and $D(a)$ is the growth rate of
fluctuations in linear theory.  Measurements of two-point clustering
as a function of angle with respect to the LOS direction can directly
constrain $f$ times the normalization of matter fluctuations
\citep[e.g.][]{2008Natur.451..541G, 2009MNRAS.393..297P,
  2009MNRAS.397.1348W}.

In this paper, we will focus on the prediction of the two-point
correlation function:
\begin{equation}
  \xi(\vec{r}) = \langle \delta(\vec{x}) \delta(\vec{x}+\vec{r})
  \rangle. 
\end{equation}
In real-space, $\xi(\vec{r}) \equiv \xi(r)$ is only a function of the
separation length, while in redshift-space $\xi(\vec{s})$ depends on
the cosine of the angle between the pair separation vector and the
LOS, $\mu_s \equiv \hat{\vec{s}} \cdot \hat{\vec{z}}$.  It is
convenient and common to condense the information in $\xi(\vec{s})$
into Legendre polynomial moments (we use $L_\ell$ for $\ell$th order
Legendre polynomial to avoid ambiguity):
\begin{equation}
  \label{eq:legendre-expansion}
  \xi(s,\mu_s) = \sum_{\ell} \xi_{\ell}(s) L_{\ell}(\mu)\ .
\end{equation}
By symmetry, odd $\ell$ moments vanish.  In linear theory, only $\ell=
0,2,4$ contribute; we will focus our model predictions on those
moments.  In \S\ref{sec:redshift-space-stat} we shall also look at
clustering ``wedges'' \citep[e.g.][]{2012MNRAS.419.3223K}, but these
require no further formalism.

Throughout this paper, we adopt the Einstein summation convention and
the following convention of Fourier transform and its inverse ($n$ is
the number of dimensions, here usually 1 or 3):
\begin{equation}
  \label{eq:fourier-def}
  \tilde{F}({\bf k}) = \int \d^nx\ F({\bf x})
  \e{-i{\bf k} \cdot {\bf x}}\ ,\quad F({\bf x}) = \int
  \dfrac{\d^nk}{(2\pi)^n}\ \tilde{F}({\bf k}) \e{i{\bf k} \cdot {\bf
      x}}\ .
\end{equation}

\subsection{Integrated Perturbation Theory (iPT): formalism for biased
tracers}

Lagrangian perturbation theories perform a perturbative expansion in
the displacement field $\vec{\Psi} = \vec{\Psi}^{(1)} +
\vec{\Psi}^{(2)} + \vec{\Psi}^{(3)} + \cdots$.  Here, $\vec{\Psi}$
relates the Eulerian (final) coordinates $\vec{x}$ and Lagrangian
(initial) coordinates $\vec{q}$ of a mass element or discrete tracer
object:
\begin{equation}
  \label{eq:lagriangian-disp-def}
  \vec{x}(\vec{q}, t) = \vec{q} + \vec{\Psi}(\vec{q},t)\ .
\end{equation}
The relation between Eulerian and Lagrangian fields of
matter density contrast ($\delta = \rho/\bar{\rho}-1$) is given by
\begin{equation}
  \label{eq:density-field-eulerian-lagrangian}
  [1+\delta_m(\vec{x},t)]\ \d^3x = [1+\delta_m(\vec{q},0)]\ \d^3q = \d^3q\ .
\end{equation}
\citet{2008PhRvD..78h3519M,2008PhRvD..77f3530M} laid
out the formalism for including redshift-space distortions and
non-linear local Lagrangian biasing within Lagrangian perturbation
theory.  The density contrast of our tracer field in Lagrangian space,
$\delta(\vec{q})$, is related to the underlying Lagrangian matter
density fluctuations smoothed on scale $R$:
\begin{equation}
  1 + \delta(\vec{q}) = F[\delta_{m,R}(\vec{q})]\ ,
\end{equation}
where $F(\delta)$ is the bias function. Note that the
smoothing scale $R$ naturally drops out in the final predictions for
all statistics of interest in this paper, but is necessary to keep
intermediate quantities well-behaved.  Thus
\begin{equation}
  \label{eq:density-field-integration}
  1 + \delta(\vec{x},t) = \int
  \d^3q\ F[\delta_{m,R}(\vec{q})] \delta^D[\vec{x} - \vec{q} -
  \vec{\Psi}(\vec{q}, t)]\ .
\end{equation}
After a coordinate transformation $\{\vec{q}_1,\vec{q}_2\} \rightarrow
\{\vec{q} = \vec{q}_2-\vec{q}_1, \mathbf{Q} = (\vec{q}_1+\vec{q}_2) /
2\}$, and expressing both $F$ and $\delta^D$ in
Eq.~\eqref{eq:density-field-integration} by their Fourier
representations, the two-point correlation function in real-space is
given by [see also equations (15) through (20) in
\citet{2012arXiv1209.0780C}]:
\begin{equation}
  \label{eq:xi-fourier}
  \begin{split}
    1 + \xi(\vec{r}) & = \int \d^3q \int \dfrac{\d^3k}{(2\pi)^3}
    \e{i\vec{k}\cdot(\vec{q}-\vec{r})} \int \dfrac{\d \lambda_1}{2\pi}
    \dfrac{\d \lambda_2}{2\pi}
    \\
    & \times\tilde{F}(\lambda_1) \tilde{F}(\lambda_2) \left\langle
      \e{i\left(\lambda_1\delta_1 + \lambda_2\delta_2 + \vec{k}\cdot
          \vec{\Delta} \right)} \right\rangle\ ,
  \end{split}
\end{equation}
where $\delta_{1,2} = \delta(\vec{q}_{1,2})$,
$\vec{\Delta}=\vec{\Psi}(\vec{q}_2) - \vec{\Psi}(\vec{q}_1)$, and
$\tilde{F}(\lambda)$ is the Fourier transform of $F(\delta_R)$ with
coordinates pair $\delta_R$ versus $\lambda$.

\section{Clustering and velocity statistics in CLPT}
\label{sec:v-sigma}

In this section we extend the work of \citet{2008PhRvD..78h3519M} and
\citet{2012arXiv1209.0780C} to enable the calculation of moments of
the pairwise velocity distribution for tracers that are biased in a
local Lagrangian sense.  Once the real-space two-point correlation
function and pairwise velocity distributions are known, they determine
the observed two-point clustering in redshift-space.  These are also
the key ingredients of the Gaussian streaming model, as we shall
discuss further in \S\ref{sec:GSM}.

\subsection{Velocity moments in CLPT: formalism}

The relative peculiar velocity between two tracers at Eulerian
coordinates $\vec{x}_1$ and $\vec{x}_2$ can be simply expressed in
terms of the time derivative of the displacement field $\vec{\Psi}$:
\begin{equation}
  \vec{u}(\vec{x}_2) - \vec{u}(\vec{x}_1) = a \left(
    \dot{\vec{x}}_2 - \dot{\vec{x}}_1\right) = a\dot{\vec{\Delta}}.
\end{equation}
In the time-independent approximation to the perturbative kernels
where ${\vec{\Psi}}^{(k)} \propto D^k$ for $D$ the linear growth
function \citep[equation (46) of][]{2008PhRvD..77f3530M},
\begin{equation}
  \dot{\vec{\Psi}}^{(k)} = kHf\vec{\Psi}^{(k)}\ .
\end{equation}
Thus we have a perturbative expansion for the Cartesian components of
$\vec{v}_n$ (adopting the units of Eq.~\ref{eq:sx}) in terms of the
components of $\vec{\Delta}_n$
\begin{equation}
  \vec{v}_n(\vec{x}_2) - \vec{v}_n(\vec{x}_1) = \sum_k
  kf \vec{\Delta}_n^{(k)} = \frac{\dot{\vec{\Delta}}_n}{H}\ .
\end{equation}
We follow common practice and define the velocity generating function
$Z(\vec{r}, \vec{J})$ by
\begin{equation}
  \label{eq:Z}
  \begin{split}
    Z(\vec{r}, \vec{J}) & = \int \d^3q \int \dfrac{\d^3k}{(2\pi)^3}
    \e{i\vec{k}\cdot(\vec{q}-\vec{r})} \int \dfrac{\d \lambda_1}{2\pi}
    \dfrac{\d \lambda_2}{2\pi}
    \\
    & \times\tilde{F}(\lambda_1) \tilde{F}(\lambda_2) \left\langle
      \e{i\left(\lambda_1\delta_1 + \lambda_2\delta_2 + \vec{k}\cdot
          \vec{\Delta} + \vec{J}\cdot\dot{\vec{\Delta}}/H\right)}
    \right\rangle\ .
  \end{split}
\end{equation}
Note that $\xi(\vec{r}) = Z(\vec{r}, 0) - 1$
(Eq.~\eqref{eq:xi-fourier}), and derivatives of $Z$ give the pairwise
velocity moments of interest (i.e., the numerators in our
Eqs.~\ref{eq:v-infall-def} and \ref{eq:vdisp-sigma-mu}):
\begin{equation}
\label{eq:vLPT}
  \begin{split}
    & \left\langle [1+\delta(\vec{x})] [1+ \delta(\vec{x}+\vec{r})]
      \left\{\prod_{k=1}^{p} [\vec{v}_{i_k}(\vec{x}+\vec{r}) - \vec{
          v}_{i_k}(\vec{x})]\right\} \right\rangle
    \\
    = & \int \d^3q \int \dfrac{\d^3k}{(2\pi)^3}
    \e{i\vec{k}\cdot(\vec{q}-\vec{r})} \int \dfrac{\d \lambda_1}{2\pi}
    \dfrac{\d \lambda_2}{2\pi}
    \\
    \times & \tilde{F}(\lambda_1) \tilde{F}(\lambda_2) \left\langle
      \left(\prod_{k=1}^{p}
        \left(\dfrac{\dot{\vec{\Delta}}_{i_k}}{H}\right)\right)
      \e{i\left(\lambda_1\delta_1 + \lambda_2\delta_2 + \vec{k}\cdot
          \vec{\Delta} \right)} \right\rangle
    \\
    = & \left.\prod_{k=1}^{p} \left(-i\dfrac{\partial}{\partial
          \vec{J}_{i_k}}\right) Z(\vec{r}, \vec{J}) \;
    \right|_{\vec{J} = 0}
  \end{split}
\end{equation}
Here the set $\{i_k\}$ specifies the Cartesian coordinate direction
for each derivative with respect to $J_{i_k}$.  Before proceeding to
evaluate Eq.~\eqref{eq:vLPT}, we generalize the definitions of the
functions $K$, $L$, and $M$ in \citet{2012arXiv1209.0780C} to include
$\vec{J}$.  These functions are convenient shorthand for intermediate
results.
\begin{equation}
  \label{eq:perturbation-notation-short}
  \begin{split}
    & X = \lambda_1\delta_1 + \lambda_2\delta_2 + \vec{k}\cdot
    \vec{\Delta} + \vec{J} \cdot \dot{\vec{\Delta}} / H\ ,
    \\
    & K_{p,\{i_1..i_p\}}(\vec{q}, \vec{k}, \lambda_1, \lambda_2) =
    \left.\left\langle \left(-i\dfrac{\partial}{\partial
            \vec{J}_{i_k}}\right)^p \e{iX} \right\rangle\
    \right|_{\vec{J}=0}\ ,
    \\
    & L_{p,\{i_1..i_p\}}(\vec{q}, \vec{k}) = \int \dfrac{\d
      \lambda_1}{2\pi} \dfrac{\d \lambda_2}{2\pi}
    K_{p,\{i_1..i_p\}}(\vec{q}, \vec{k}, \lambda_1, \lambda_2)\ ,
    \\
    & M_{p,\{i_1..i_p\}}(\vec{r}, \vec{q}) = \int
    \dfrac{\d^3k}{(2\pi)^3} \e{i\vec{k}\cdot(\vec{q}-\vec{r})}
    L_{p,\{i_1..i_p\}}(\vec{q}, \vec{k})\ ,
    \\
    & \left\langle [1+\delta(\vec{x})] [1+ \delta(\vec{x}+\vec{r})]
      \left\{\prod_{k=1}^{p} [\vec{v}_{i_k}(\vec{x}+\vec{r}) -
        \vec{v}_{i_k} (\vec{x})]\right\} \right\rangle
    \\
    & \quad =\int \d^3q M_{p,\{i_1,..,i_p\}}(\vec{r}, \vec{q})\ .
  \end{split}
\end{equation}
The first subscript of these functions indicates the
number of derivative terms $p$, and the second is a list containing
the Cartesian indices of the $p$ derivatives.

\section{Evaluating the CLPT predictions}
\label{sec:evaluate}
\subsection{Evaluating $\xi(r)$ in CLPT}

To begin, we review the calculation of the real-space two-point
correlation function $\xi(r)$ in CLPT, first presented in
\citet{2012arXiv1209.0780C}, building upon the work of
\citet{2008PhRvD..78h3519M,2008PhRvD..77f3530M}.  The cumulant
expansion theorem,
\begin{equation}
  \label{eq:cumulant-expansion}
  \left\langle \e{iX} \right\rangle = \exp\left[ \sum_{N=1}^\infty
    \dfrac{i^N}{N!} \langle X^N \rangle_c \right]\ ,
\end{equation} makes the evaluation of $K$ tractable; here $\langle
X^N \rangle_c$ is the $N$th cumulant of the random variable $X$.
Taylor expanding the exponential on the right hand side of
Eq.~\eqref{eq:cumulant-expansion} only for terms that vanish in the
limit $|\vec{q}| \rightarrow \infty$ and keeping only terms up to
$O(P_L^2)$, \citet{2012arXiv1209.0780C}, Sec.~4 obtain
\begin{equation}
  \label{eq:xi-kernel-k}
  \begin{split}
    & K_0 = \left. \left\langle \e{iX} \right\rangle
    \right|_{\vec{J}=0}
    \\
    & = \e{-(1/2)A_{ij}k_ik_j} \e{-(1/2)(\lambda_1^2+\lambda_2^2)
      \sigma_R^2} \bigg\{ 1 - \lambda_1\lambda_2 \xi_L +
    \dfrac{1}{2}\lambda_1^2\lambda_2^2 \xi_L^2
    \\
    & - (\lambda_1+\lambda_2) U_ik_i + \dfrac{1}{2}
    (\lambda_1+\lambda_2)^2 U_iU_jk_ik_j - \dfrac{i}{6}
    W_{ijk}k_ik_jk_k
    \\
    & + \lambda_1\lambda_2(\lambda_1+\lambda_2) \xi_L U_ik_i -
    \dfrac{i}{2} (\lambda_1+\lambda_2) A^{10}_{ij}k_ik_j
    \\
    & + \dfrac{i}{2}(\lambda_1^2 +\lambda_2^2) U^{20}_ik_i -
    i\lambda_1\lambda_2U^{11}_ik_i + O(P_L^3) \bigg\}\ .
  \end{split}
\end{equation}
In Eq.~\eqref{eq:xi-kernel-k}, we adopt the following short-hand
definitions:
\begin{equation}
  \label{eq:xi-short-def}
  \begin{split}
    & \langle \delta_1^2 \rangle_c = \langle \delta_2^2 \rangle_c =
    \sigma_R^2\ ,\quad \langle \delta_1\delta_2 \rangle_c =
    \xi_L(\vec{q})\ ,
    \\
    & U^{mn}_i = \langle \delta_1^m \delta_2^n \Delta_i \rangle_c\ ,
    \quad A^{mn}_{ij} = \langle \delta_1^m \delta_2^n \Delta_i\Delta_j
    \rangle_c\ ,
    \\
    & W^{mn}_{ijk} = \langle \delta_1^m \delta_2^n
    \Delta_i\Delta_j\Delta_k \rangle_c\ ,
    \\
    & U_i = U^{10}_i\ ,\quad A_{ij} = A^{00}_{ij}\ ,\quad W_{ijk} =
    W^{00}_{ijk}\ .
  \end{split}
\end{equation}
Subsequently we apply
\citep[see][]{2008PhRvD..78h3519M,2012arXiv1209.0780C}
\begin{equation}
  \label{eq:bias-expect-identity}
  \int \dfrac{\d \lambda}{2\pi} \tilde{F}(\lambda)
  \e{\lambda^2\sigma_R^2/2} (i\lambda)^n = \langle F^{(n)} \rangle\ ,
\end{equation}
where $\langle F^{(n)} \rangle$ is the expectation value of the $n$th
derivative of $F(\delta_R)$. This relation enables us to conduct such
transformations from integration with respect to $\lambda$ to bias
parameters:
\begin{equation}
  \label{eq:bias-transform}
  \begin{split}
    & (\lambda_1 + \lambda_2) \rightarrow 2\fp{}\ , \qquad
    (\lambda_1^2 + \lambda_2^2 ) \rightarrow 2\fpp{}\ ,
    \\
    & \lambda_1 \lambda_2 \rightarrow \fp{}^2\ , \qquad \lambda_1^2
    \lambda_2^2\rightarrow \fpp{}^2\ ,
    \\
    & \lambda_1 \lambda_2 (\lambda_1 + \lambda_2) \rightarrow
    2\fp{}\fpp{}\ .
  \end{split}
\end{equation}
We can hence evaluate $L_0$ analytically:
\begin{equation}
  \label{eq:xi-kernel-l}
  \begin{split}
    & L_0 = \e{-(1/2)A_{ij}k_ik_j} \bigg\{ 1 + \fp{}^2 \xi_L +
    \dfrac{1}{2}\fpp{}^2 \xi_L^2 + 2i \fp{} U_ik_i
    \\
    & - [\fpp{} + \fp{}^2] U_iU_jk_ik_j - \dfrac{i}{6}
    W_{ijk}k_ik_jk_k - \fp{} A^{10}_{ij}k_ik_j
    \\
    & + 2i \fp{} \fpp{} \xi_L U_ik_i + i\fpp{} U^{20}_ik_i
    \\
    & - i\fp{}^2U^{11}_ik_i + O(P_L^3) \bigg\}\ .
  \end{split}
\end{equation}
The integration with respect to $\vec{k}$ is then conducted, using
some basic relations for Gaussian integration \citep[see appendix C
of][] {2012arXiv1209.0780C}, which gives
\begin{equation}
  \label{eq:xi-kernel-m}
  \begin{split}
    & M_0 = \int \dfrac{\d^3 k}{(2\pi)^3}
    \e{i\vec{k}\cdot(\vec{q}-\vec{r})} L_0(\vec{q},\vec{k})
    \\
    & = \dfrac{1}{(2\pi)^{3/2}|A|^{1/2}}\e{-(1/2)(A^{-1})_{ij}
      (q_i-r_i)  (q_j-r_j)}
    \\
    &\times \bigg\{ 1 + \fp{}^2 \xi_L + \dfrac{1}{2}\fpp{}^2
    \xi_L^2 - 2 \fp{} U_ig_i + \dfrac{1}{6} W_{ijk}\Gamma_{ijk}
    \\
    & - [\fpp{} + \fp{}^2] U_iU_jG_{ij} - \fp{}^2U^{11}_ig_i - \fpp{}
    U^{20}_ig_i
    \\
    & - 2 \fp{} \fpp{} \xi_L U_ig_i - \fp{} A^{10}_{ij}G_{ij} +
    O(P_L^3) \bigg\}\ .
  \end{split}
\end{equation}
Here we define
\begin{equation}
  \label{eq:gaussian-intg-aux}
  \begin{split}
    & g_i = (A^{-1})_{ij}(q_j-r_j)\ ,\quad G_{ij} = (A^{-1})_{ij} -
    g_ig_j\ ,
    \\
    & \Gamma_{ijk} = (A^{-1})_{ij} g_k + (A^{-1})_{ki} g_j +
    (A^{-1})_{jk} g_i - g_ig_jg_k\ .
  \end{split}
\end{equation}
Finally, the desired correlation function is given by
the $\vec{q}$ integration of $M_0$:
\begin{equation}
  \label{eq:xi-exp} 1 + \xi(\vec{r}) = \int \d^3q\
  M_0(\vec{r},\vec{q})\ .
\end{equation}
In order to evaluate $M_0$, we expand $U^{mn}_{i}$, $A^{mn}_{ij}$ and
$W^{mn}_{ijk}$ in \eqref{eq:xi-short-def} with respect to
$\vec{\Delta}^{(n)}$, i.e.
\begin{equation}
  \label{eq:xi-correlator-different-order}
  \begin{split}
    & U^{mn(p)}_i = \langle \delta_1^m \delta_2^n \Delta_i^{(p)}
    \rangle_c\ , \quad A^{mn(pq)}_{ij} = \langle \delta_1^m \delta_2^n
    \Delta_i^{(p)} \Delta_j^{(q)} \rangle_c\ ,
    \\
    & W^{mn(pqr)}_{ijk} = \langle \delta_1^m \delta_2^n \Delta_i^{(p)}
    \Delta_j^{(q)} \Delta_k^{(r)} \rangle_c\ ,
  \end{split}
\end{equation}
and up to desired order, we have
\begin{equation}
  \label{eq:xi-correlator-desired-order}
  \begin{split}
    & U_i = U^{(1)}_i + U^{(3)}_i + \cdots\ ,
    \\
    & U^{20}_i = U^{20(2)}_i + \cdots\ ,\quad U^{11}_i = U^{11(2)}_i +
    \cdots\ ,
    \\
    & A_{ij} = A^{(11)}_{ij} + A^{(22)}_{ij} + A^{(13)}_{ij} +
    A^{(31)}_{ij} + \cdots\ ,
    \\
    & A^{10}_{ij} = A^{10(12)}_{ij} + A^{10(21)}_{ij} + \cdots\ ,
    \\
    & W_{ijk} = W^{(112)}_{ijk} + W^{(121)}_{ijk} + W^{(211)}_{ijk} +
    \cdots \ .
  \end{split}
\end{equation} We refer the readers to appendices B through C in
\cite{2012arXiv1209.0780C} for details of evaluating those
correlators.

\subsection{The mean pairwise velocity in CLPT}

To compute the mean pairwise velocity in CLPT, we first evaluate
$K_{1,n}$, again making use of the cumulant expansion theorem:
\begin{equation}
  \label{eq:v-kernel-k-full}
  \begin{split}
    & K_{1,n}(\lambda_1, \lambda_2, \vec{k}, \vec{q}) =
    \\
    & = \left. \exp{ \left[ \sum_{N=0}^\infty\dfrac{i^N}{N!}
          \left\langle X^N \right\rangle_c \right] } \left[
        \sum_{N=0}^\infty\dfrac{i^N}{N!}  \left\langle
          \dfrac{\dot{\Delta}_n}{H} X^N \right\rangle_c \right]\
    \right|_{\vec{J}=0} \ .
  \end{split}
\end{equation} Up to the second order of the linear power spectrum
[i.e. $O(P_L^2)$], Eq.~\eqref{eq:v-kernel-k-full} is recast as
\begin{equation}
  \label{eq:v-kernel-k-2nd}
  \begin{split}
    & K_{1,n}(\lambda_1, \lambda_2, \vec{k}, \vec{q})
    \\
    & = f \e{-(1/2)A_{ij}k_ik_j}
    \e{-(1/2)(\lambda_1^2+\lambda_2^2)\sigma_R^2}
    \\
    & \times \bigg\{ i(\lambda_1+\lambda_2)\dot{U}_n + i k_i
    \dot{A}_{in} - \dfrac{1}{2}(\lambda_1^2+\lambda_2^2)
    \dot{U}^{20}_n
    \\
    & - \lambda_1\lambda_2 \dot{U}^{11}_n - \dfrac{1}{2}
    k_ik_j\dot{W}_{ijn} -(\lambda_1+\lambda_2) k_i \dot{A}^{10}_{in}
    \\
    & - i\lambda_1\lambda_2(\lambda_1+\lambda_2) \xi_L \dot{U}_n
    -i(\lambda_1+\lambda_2)^2 k_i U_i \dot{U}_n
    \\
    & - i\lambda_1\lambda_2 \xi_L k_i\dot{A}_{in} -
    i(\lambda_1+\lambda_2) k_ik_jU_i\dot{A}_{in} + O(P_L^3) \bigg\}\ .
  \end{split}
\end{equation}
where we define (up to desired order)
\begin{equation}
  \label{eq:vd-sym-def}
  \begin{split}
    & \dot{U}_n = \dfrac{\langle \delta_1 \dot{\Delta}_n \rangle}{f} =
    U^{(1)}_n + 3 U^{(3)}_n + \cdots\ ,
    \\
    & \dot{U}^{20}_n = \dfrac{\langle \delta_1^2 \dot{\Delta}_n
      \rangle}{f} = U^{20(2)}_n + \cdots\ ,
    \\
    & \dot{U}^{11}_n = \dfrac{\langle \delta_1 \delta_2 \dot{\Delta}_n
      \rangle}{f} = U^{11(2)}_n + \cdots\ ,
    \\
    & \dot{A}_{in} = \dfrac{\langle \Delta_i \dot{\Delta}_n
      \rangle}{f} = A^{(11)}_{in} + 3A^{(13)}_{in} + A^{(31)}_{in} +
    2A^{(22)}_{in} + \cdots\ ,
    \\
    & \dot{A}^{10}_{in} = \dfrac{\langle \delta_1 \Delta_i
      \dot{\Delta}_n \rangle}{f} = 2A^{10(12)}_{in} +
    A^{10(21)}_{in}+\cdots\ ,
    \\
    & \dot{W}_{ijn} = \dfrac{\langle \delta_1 \Delta_i \Delta_j
      \dot{\Delta}_n \rangle}{f} = 2 W^{(112)}_{ijn} + W^{(121)}_{ijn}
    + W^{(211)}_{ijn}+\cdots\ .
  \end{split}
\end{equation}
Integrate with respect to $\lambda_1$ and $\lambda_2$, we have
\citep[see the appendices of][]{2012arXiv1209.0780C}
\begin{equation}
  \label{eq:v-kernel-l-2nd}
  \begin{split}
    & L_{1,n} = \int \dfrac{\d \lambda_1}{2\pi} \dfrac{\d
      \lambda_2}{2\pi} \tilde{F}(\lambda_1) \tilde{F}(\lambda_2)
    K_{1,n}(\lambda_1, \lambda_2, \vec{k}, \vec{q})
    \\
    & = f \e{-(1/2)A_{ij}k_ik_j}
    \e{-(1/2)(\lambda_1^2+\lambda_2^2)\sigma_R^2}
    \\
    & \times \bigg\{ 2 \fp{}\dot{U}_n + i k_i \dot{A}_{in} + \fpp{}
    \dot{U}^{20}_n + \fp{}^2 \dot{U}^{11}_n
    \\
    & - \dfrac{1}{2} k_ik_j\dot{W}_{ijn} + 2i \fp{} k_i
    \dot{A}^{10}_{in} + 2\fp{} \fpp{} \xi_L \dot{U}_n
    \\
    & + 2i[\fpp{}+\fp{}^2] k_i U_i \dot{U}_n + i \fp{}^2 \xi_L
    k_i\dot{A}_{in}
    \\
    & - 2\fp{} k_ik_jU_i\dot{A}_{in} + O(P_L^3) \bigg\}\ .
  \end{split}
\end{equation}
Then evaluate the integration over $\vec{k}$, we have
\begin{equation}
  \label{eq:v-kernel-m-2nd}
  \begin{split}
    & M_{1,n} = \int \dfrac{\d^3 k}{(2\pi)^3}
    \e{i\vec{k}\cdot(\vec{q}-\vec{r})} L_{1,n}(\vec{q},\vec{k})
    \\
    & = \dfrac{f^2}{(2\pi)^{3/2}|A|^{1/2}}
    \e{-(1/2)(A^{-1})_{ij}(q_i-r_i)(q_j-r_j)}
    \\
    & \times \bigg\{ 2 \fp{}\dot{U}_n - g_i\dot{A}_{in} + \fpp{}
    \dot{U}^{20}_n + \fp{}^2 \dot{U}^{11_n}
    \\
    & - \dfrac{1}{2} G_{ij}\dot{W}_{ijn} - 2 \fp{} g_i
    \dot{A}^{10}_{in} + 2\fp{} \fpp{} \xi_L \dot{U}_n
    \\
    & - 2 [\fpp{}+\fp{}^2] g_i U_i \dot{U}_n - \fp{}^2 \xi_L
    g_i\dot{A}_{in}
    \\
    & - 2\fp{} G_{ij}U_i\dot{A}_{in} + O(P_L^3) \bigg\}\ ,
  \end{split}
\end{equation}
and finally,
\begin{equation}
  \label{eq:v-exp} v_{12,n}(\vec{r}) = [1+\xi(r)]^{-1} \int \d^3q\
M_{1,n}(\vec{r}, \vec{q})\ .
\end{equation}
Typically $v_{12,n}$ is projected along the direction
of pair separation vector, i.e. $v_{12} = v_{12,n}\hat{r}_n$.

\subsection{The pairwise velocity dispersion in CLPT}

The integration kernel for the velocity dispersion tensor is
\begin{equation}
  \label{eq:vd-kernel-k-full}
  \begin{split}
    & K_{2,nm}(\lambda_1, \lambda_2, \vec{k}, \vec{q}) = \left. \left(
        -i\dfrac{\partial}{\partial J_m}\right) \left(
        -i\dfrac{\partial}{\partial J_n}\right) \left\langle
        \e{iX(\vec{J})}\right\rangle \right|_{\vec{J}\rightarrow 0}
    \\
    & = \exp{ \left[ \sum_{N=0}^\infty\dfrac{i^N}{N!}  \left\langle
          X^N \right\rangle_c \right] } \Bigg\{ \left[
      \sum_{N=0}^\infty\dfrac{i^N}{N!}  \left\langle \dot{\Delta}_n
        \dot{\Delta}_m X^N \right\rangle_c \right]
    \\
    & \quad + \left[ \sum_{N=0}^\infty\dfrac{i^N}{N!} \left\langle
        \dot{\Delta}_n X^N \right\rangle_c \right] \left[
      \sum_{M=0}^\infty\dfrac{i^M}{M!}  \left\langle \dot{\Delta}_m
        X^M \right\rangle_c \right] \Bigg\}\ ,
  \end{split}
\end{equation}
Expanding Eq.~\eqref{eq:vd-kernel-k-full} to second order in the
linear power spectrum (i.e.~to the order of $O(P_L^2)$):
\begin{equation}
  \label{eq:vd-kerner-k-2nd}
  \begin{split} &
    K_{2,nm}(\lambda_1, \lambda_2, \vec{k}, \vec{q})
    \\
    & = f^2 \e{-(1/2)A_{ij}k_ik_j}
    \e{-(1/2)(\lambda_1^2+\lambda_2^2)\sigma_R^2}
    \\
    & \times \big\{ (\lambda_1+\lambda_2)^2 \dot{U}_n \dot{U}_m -
    (\lambda_1+\lambda_2) ( \dot{A}_{in}k_i\dot{U}_m +
    \dot{A}_{im}k_i\dot{U}_n) 
    \\
    & - \dot{A}_{im}k_i\dot{A}_{jn}k_j + [1 -\lambda_1\lambda_2 \xi_L -
    (\lambda_1+\lambda_2) U_ik_i ] \ddot{A}_{nm}
    \\
    & + i (\lambda_1+\lambda_2) \ddot{A}^{10}_{nm} + i
    \ddot{W}_{inm}k_i + O(P_L^3) \big\}\ ,
  \end{split}
\end{equation}
where we define (up to the desired order)
\begin{equation}
  \label{eq:vd-sym-def}
  \begin{split}
    & \ddot{A}_{nm} = \dfrac{\langle \dot{\Delta}_n\dot{\Delta}_m
      \rangle}{f^2} = A^{(11)}_{nm} + 3 A^{(13)}_{nm} + 3
    A^{(31)}_{nm} + 4 A^{(22)}_{nm}\ ,
    \\
    & \ddot{A}_{10,nm} = \dfrac{\langle \delta_1
      \dot{\Delta}_n\dot{\Delta}_m \rangle}{f^2} = 2A^{10(12)}_{nm} +
    2A^{10(21)}_{nm}\ ,
    \\
    & \ddot{W}_{inm} = \dfrac{\langle \delta_1 \Delta_i
      \dot{\Delta}_n\dot{\Delta}_m \rangle}{f^2} = 2 W^{(112)}_{inm} +
    2 W^{(121)}_{inm} + W^{(211)}_{inm}\ .
  \end{split}
\end{equation}
Then evaluate the integration with respect to $\lambda_1$,
$\lambda_2$, we have
\begin{equation}
  \label{eq:vd-kernel-l-2nd}
  \begin{split}
    & L_{2,nm} = \int \dfrac{\d \lambda_1}{(2\pi)} \dfrac{\d
      \lambda_2}{(2\pi)} \tilde{F}(\lambda_1) \tilde{F}(\lambda_2)
    K_{2,nm}(\lambda_1, \lambda_2, \vec{k}, \vec{q})
    \\
    & = \big\{ 2 [\fp{}^2 + \fpp{} ] \dot{U}_n \dot{U}_m + 2 i \fp{} (
    \dot{A}_{in}k_i\dot{U}_m + \dot{A}_{im}k_i\dot{U}_n)
    \\
    & - \dot{A}_{im}k_i\dot{A}_{jn}k_j + [1 + \fp{}^2 \xi_L + 2i \fp{}
    U_ik_i ] \ddot{A}_{nm}
    \\
    & + 2\fp{} \ddot{A}^{10}_{nm} + i \ddot{W}_{inm}k_i + O(P_L^3)
    \big\} f^2 \e{-(1/2)A_{ij}k_ik_j}\ ,
  \end{split}
\end{equation}
and then with respect to $\vec{k}$:
\begin{equation}
  \label{eq:vd-kernel-m-2nd}
  \begin{split}
    & M_{2,nm} = \int \dfrac{\d^3 k}{(2\pi)^3}
    \e{i\vec{k}\cdot(\vec{q}-\vec{r})} L_{2,nm}(\vec{q},\vec{k})
    \\
    & = \dfrac{f^2}{(2\pi)^{3/2}|A|^{1/2}}
    \e{-(1/2)(A^{-1})_{ij}(q_i-r_i)(q_j-r_j)}
    \\
    & \times \big\{ 2 [\fp{}^2 + \fpp{} ] \dot{U}_n \dot{U}_m - 2
    \fp{} ( \dot{A}_{in}g_i\dot{U}_m + \dot{A}_{im}g_i\dot{U}_n)
    \\
    & - \dot{A}_{im}\dot{A}_{jn}G_{ij} + [1 + \fp{}^2 \xi_L - 2\fp{}
    U_ig_i ] \ddot{A}_{nm}
    \\
    & + 2\fp{} \ddot{A}^{10}_{nm} - \ddot{W}_{inm}g_i + O(P_L^3)
    \big\}\ .
  \end{split}
\end{equation}
Finally, $\vd{12,nm}$ can be obtained by
\begin{equation}
  \label{eq:vd-exp}
  \vd{12,nm}(\vec{r}) = [1+\xi(r)]^{-1} \int \d^3q\
  M_{2,nm}(\vec{r}, \vec{q})\ .
\end{equation}
Desired component of pairwise velocity dispersion can be obtained by
different components or kinds of contractions of the tensor
$\vd{12,nm}$. In order to obtain the velocity dispersion components
parallel to and perpendicular to the pairwise separation unit vector
$\hat{r}$, we project $\vd{12,nm}$ into different directions:
\begin{equation}
  \label{eq:vd-components}
  \vd{\parallel} = \vd{12,nm} \hat{r}_n
  \hat{r}_m\ ,\quad \vd{\bot} = ( \vd{12,nm} \delta^K_{nm} -
  \vd{\parallel} ) / 2\ .
\end{equation}

\subsection{Cross-correlation of halos with different bias parameters}
\label{sec:cross-corr-theory}

It is a straightforward generalization of the above to handle
cross-correlations between two tracers with different biases.  We note
that the displacement field $\vec{\Psi}$ is identical for all species
-- the difference is only in their bias parameters.  Therefore, in
this ``cross-correlation'' scenario we have different $\tilde{F_j}$
($j=1$ or $2$) for $\lambda_1$ and $\lambda_2$. Equation
\eqref{eq:bias-expect-identity} is hence recast as:
\begin{equation}
  \label{eq:bias-expect-identity-different-species}
  \int \dfrac{\d \lambda_j}{2\pi} \tilde{F_j}(\lambda_j)
  \e{\lambda_j^2\sigma_R^2/2} (i\lambda_j)^n = \langle F_j^{(n)}
  \rangle\ ,\quad j = 1,\ 2\ . 
\end{equation}
Hence we can adopt a list of transformations for bias
parameters to obtain cross-correlation between different species,
which can be straightforwardly deduced from
Eq.~\eqref{eq:bias-expect-identity-different-species}:
\begin{equation}
  \label{eq:bias-transform-cross-correlation}
  \begin{split}
    & (\lambda_1 + \lambda_2) \rightarrow [\fp{1} + \fp{2}]\ ,
    \\
    & (\lambda_1^2 + \lambda_2^2 ) \rightarrow [\fpp{1} + \fpp{2}]\ ,
    \\
    & \lambda_1 \lambda_2 \rightarrow \fp{1} \fp{2}\ ,
    \\
    & \lambda_1^2 \lambda_2^2\rightarrow \fpp{1} \fpp{2}\ ,
    \\
    & \lambda_1 \lambda_2 (\lambda_1 + \lambda_2) \rightarrow
    [ \fp{1}\fpp{2} + \fp{2}\fpp{1} ]\ .
  \end{split}
\end{equation}
Those transformations can also be derived and verified by using the
symmetry in $\lambda_1$ and $\lambda_2$ of the relevant
expressions. They can be applied to Eqs.~\eqref{eq:xi-kernel-m},
\eqref{eq:v-kernel-m-2nd}, and \eqref{eq:vd-kernel-m-2nd}, with
$\fp{j}$ and $\fpp{j}$ ($j=1\ ,2$) being bias parameters for two
different species. It is easy to verify that the cross-correlation
expressions reduce to the auto-correlation expressions when
$\fp{1}=\fp{2}$ and $\fpp{1}=\fpp{2}$.

\section{The Gaussian streaming model}
\label{sec:GSM}

Both iPT and CLPT have difficulties reproducing the redshift space
clustering of biased tracers on small scales.  An alternative is the
``Gaussian streaming model'' introduced in
\citet{2011MNRAS.417.1913R}, which takes as inputs perturbation theory
expressions for the real-space correlation function and the velocity
statistics.

The clustering of a population of objects in redshift-space can be
related to their underlying real-space clustering and the full
pairwise velocity distribution by \citep{1995ApJ...448..494F,
  2004PhRvD..70h3007S}
\begin{equation}
  1+\xi(r_p, r_\parallel) = \int_{-\infty}^{\infty}dy\
  [1+\xi(r)] \mathcal{P}(v_z = r_\parallel - y, \vec{r}).
\end{equation}
Here $r_p$ is the transverse separation in both real and
redshift-space, $r_\parallel$ is the LOS pair separation in redshift
space, and $y$ is the LOS separation in real-space, so that $r^2 =
r_p^2 + y^2$.  \citet{2011MNRAS.417.1913R} showed that even though the
true $\mathcal{P}$ is certainly non-Gaussian, approximating it with a
Gaussian provides an accurate description of the redshift space
correlation function of massive halos:
\begin{equation}
  \label{eq:streaming-xi-s}
  \begin{split}
    1 + \xi^s(r_p, r_\parallel) & = \int \dfrac{\d y} {[2\pi
      \vd{12}(r,\mu)]^{1/2}} [1 + \xi(r)] \\ & \times \exp \left\{
      -\dfrac{[r_\parallel - y - \mu v_{12}(r)]^2}{2\vd{12}(r,\mu)}
    \right\}\ ,
  \end{split}
\end{equation}
In the scale-dependent Gaussian streaming model, the Gaussian
probability distribution function is centered at $\mu v_{12}(r)$, the
mean LOS velocity between a pair of tracers as a function of their
real space separation:
\begin{equation}
  \label{eq:v-infall-def}
  v_{12}(r) \vec{\hat{r}} = \frac{\left\langle
      [1+\delta(\vec{x})] [1+ \delta(\vec{x}+\vec{r})]
      [\vec{v}(\vec{x}+\vec{r}) - \vec{ v}(\vec{x})] \right\rangle}
  {\left\langle [1+\delta(\vec{x})] [1+ \delta(\vec{x}+\vec{r})]
    \right\rangle}
\end{equation}
The factor $[1+\delta(\vec{x})] [1+ \delta(\vec{x}+\vec{r})]$ in the
numerator and the denominator specifies that we are computing the
average relative velocity over pairs of tracers, rather than over
randomly chosen points in space.  By symmetry, the mean velocity is
directed along the pair separation vector; projecting it onto the LOS
brings a factor of $\mu = y/r$ in Eq.~\eqref{eq:streaming-xi-s}.
Similarly, the width of the velocity PDF is different for components
along and perpendicular to the pair separation vector $\hat{r}$, so
the LOS ($\hat{z}$) velocity dispersion can be decomposed as a sum
with contributions from two one-dimensional velocity dispersions.
\begin{equation}
  \label{eq:vdisp-sigma-mu}
  \begin{split}
    \vd{12}(r, \mu) & = \dfrac{\left\langle [1+\delta(\vec{x})] [1+
        \delta(\vec{x}+\vec{r})] [v_z(\vec{x}+\vec{r}) -
        v_z(\vec{x})]^2 \right\rangle}{\left\langle
        [1+\delta(\vec{x})] [1+\delta(\vec{x}+\vec{r})] \right\rangle}
    \\
    & = \mu^2 \vd{\parallel}(r) + (1-\mu^2) \vd{\bot}(r)\ .
\end{split}
\end{equation}
Linear theory expressions for $v_{12}(r)$ and $\vd{\bot,\parallel}(r)$
are given in \citet{1995ApJ...448..494F, 1988ApJ...332L...7G,
  1989ApJ...344....1G, 2011MNRAS.417.1913R}.  One finds that the
pairwise mean infall velocity, $v_{12}(r)$, is proportional to $bf$,
while $\vd{12}(r)$ scales as $f^2$ with no dependence on the
large-scale bias $b$ at linear order.

\citet{2011MNRAS.417.1913R} evaluated Eqs.~\eqref{eq:v-infall-def} and
\eqref{eq:vdisp-sigma-mu} in standard perturbation theory under the
assumption of a linear bias $b$ relating the tracer and matter density
fields, $\delta_t(\vec{x}) = b\,\delta_m(\vec{x})$.  There were several
shortcomings of this approach however.  First, standard perturbation
theory does an unsatisfactory job of describing the smoothing of the
BAO features in the real-space correlation function.  As a result, the
analysis of \citet{2012MNRAS.426.2719R} used iPT to model 
$\xi^s(r_p, r_\parallel)$
above separations of $70\,\Mpch$.  Second, the inaccuracy of the
streaming model results with standard perturbation theory inputs for
the velocity statistics can be traced to inaccuracies in the
perturbative calculation of $v_{12}$ and its derivative, $\d v_{12}/\d
r$.  This inaccuracy was smallest for halos with second-order bias
near zero, which raises the question of whether the source of
inaccuracy was the neglect of second order bias terms in the
\citet{2011MNRAS.417.1913R} calculation.  CLPT naturally includes
higher-order bias corrections, and will allow us to quantify the size
of the second order contributions to the velocity statistics of
interest.  For these reasons, we shall consider the combination of
CLPT statistics within the Gaussian streaming model ansatz.

\section{Results}
\label{sec:results-statistics}

We implemented the formulae above in a C++ code\footnote{The code is
  available at {\tt https://github.com/wll745881210/CLPT\_GSRSD.git}},
which numerically evaluates the integrations in
Eqs.~\eqref{eq:xi-exp}, \eqref{eq:v-exp} and \eqref{eq:vd-exp} for
CLPT statistics, and in Eq.~\eqref{eq:streaming-xi-s} for the Gaussian
streaming model.  In this section we present the results, and compare
them with pertinent simulation statistics.  The $N$-body simulation
set used in this work is described in more detail in
\citep{2011MNRAS.417.1913R, 2011ApJ...728..126W}, in which the
halo catalogues are constructed by FoF method.  Table
\ref{table:bias-parameters} lists the halo mass bins we use to compare
with our analytic predictions.

\subsection{Auto-correlation of halos}
\label{sec:auto-corr}

\subsubsection{Real-space auto-correlation statistics}
\label{sec:real-space-auto}

\begin{figure}
  \centering
  \includegraphics[width=80mm, keepaspectratio]{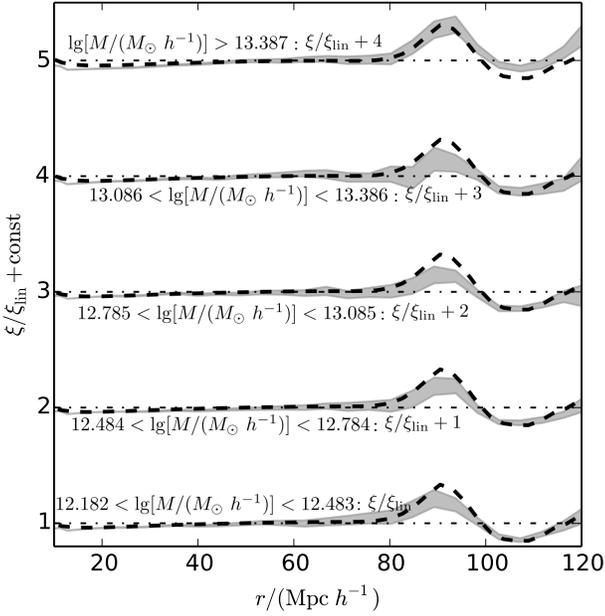}
  \caption{Real-space correlation function for halos in different mass
    bins (refer to Table \ref{table:bias-parameters}). CLPT results
    are given by heavy dashed curves; simulation results are presented
    by shaded bands showing the error range. They are divided by
    linear theory results ($\xi_\mathrm{lin}(r)$) to remove the trend,
    and different mass bins are elevated by different constants
    (labelled in the figure) to show each more clearly.  The bias
    parameters, $\fp{}$ and $\fpp{}$, for the CLPT model are given in
    Table \ref{table:bias-parameters}.}
  \label{fig:real-space-xi}
\end{figure}

\begin{figure}
  \centering
  \includegraphics[width=80mm, keepaspectratio]{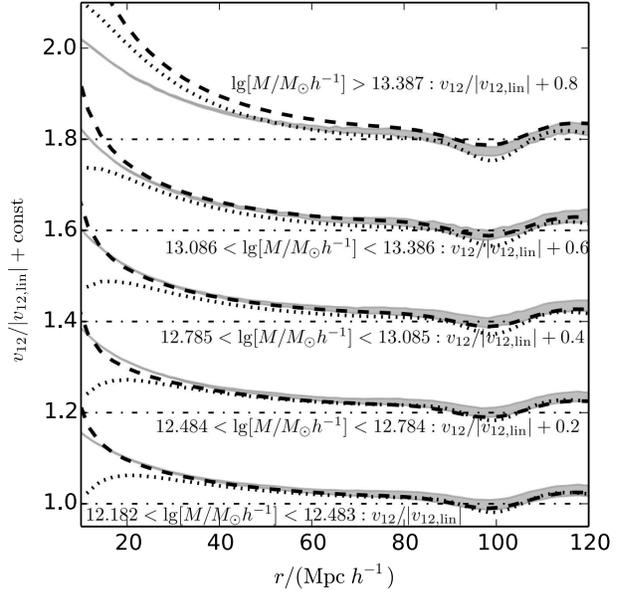}
  \caption{Pairwise infall velocity of halos in five different mass
    bins. Our CLPT results are shown by heavy dashed curves and the
    SPT results \citep{2011MNRAS.417.1913R} are shown by heavy dotted curves.
    Both are to be compared with the simulation results shown by shaded bands
    (indicating the error on the mean of the simulations).
    All results are divided by the absolute value of
    linear theory results ($|v_{12,\mathrm{lin}}(r)|$) for better
    comparison, and are elevated by different constants.}
  \label{fig:real-space-v12}
\end{figure}

\begin{figure}
  \centering
  \includegraphics[width=80mm,keepaspectratio]{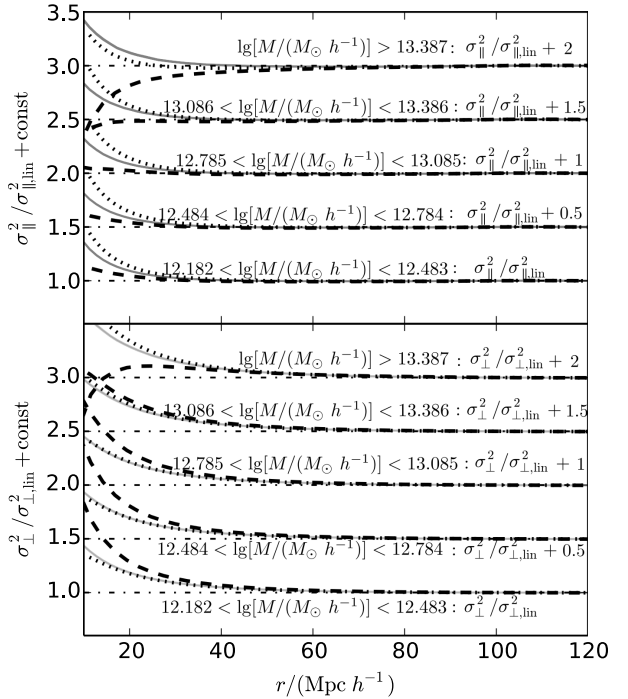}
  \caption{Pairwise velocity dispersion for halos in five different
    mass bins. For clearer presentation we show the values of
    $\vd{\parallel} / \vd{\parallel,\mathrm{lin}}$ (upper panel) and
    $\vd{\bot} / \vd{\bot,\mathrm{lin}}$ (lower panel). Shaded bands
    (very narrow) show the simulation results, the SPT results
    \citep{2011MNRAS.417.1913R} are presented by heavy dotted curves,
    and our theory prediction is presented by dashed curves. Different
    curves are elevated by different constants.}
  \label{fig:real-space-sigma12}
\end{figure}

\begin{table}
  \small
  \centering
  \caption{Bias parameters, as well as the uncertainty (denoted by
    $\sigma$) obtained by fitting the real-space correlation functions
    for different halo mass bins. Please note that $\fp{}$ and
    $\fpp{}$ are both free parameters determined by fitting and $\sigma$
    is the formal error on the fit assuming Gaussian, uncorrelated errors
    on $\xi$ as determined from the variance in the simulations.} 
  \label{table:bias-parameters}
  \vspace{0.3cm}
  \begin{tabular}{ccccc}
    \hline
    $\lg{[M/(M_\odot h^{-1})]}$ & $\fp{}$ & $\sigma(\fp{})$
    & $\fpp{}$ & $\sigma(\fpp{})$ \\
    \hline
    $12.182-12.483$ & $0.341$ & $0.004$ &  $0.06$ & $0.06$ \\
    $12.484-12.784$ & $0.435$ & $0.004$ &  $0.16$ & $0.06$ \\
    $12.785-13.085$ & $0.652$ & $0.005$ &  $0.16$ & $0.01$ \\
    $13.086-13.386$ & $0.965$ & $0.006$ &  $0.14$ & $0.09$ \\
    $> 13.387$ & $1.738$ & $0.007$ & $-0.10$ & $0.11$ \\
    \hline
  \end{tabular}
\end{table}

We present the CLPT predictions of real-space statistics in this
section, which will be used as the ``input'' of the Gaussian streaming
redshift-space distortion model. All calculations are compared with
pertinent results in Section \ref{sec:v-sigma}.

We treat $\fp{}$ and $\fpp{}$ as free parameters in our model and fit
them to the real-sapce correlation function, $\xi(r)$, measured in the
N-body simulations for each halo mass bin.  We treat all of the
$\xi(r)$ bins as independent and use the inverse variance obtained from
the simulation.  While it is incorrect to neglect the
correlations, one can see by eye that the resulting best-fit (and
the value of $\chi^2$) are entirely reasonable.  The resulting
values are listed in Table \ref{table:bias-parameters}. We note here
that it is also possible to obtain $\fpp{}$ as a function of $\fp{}$
using the peak-background split relation \citep[as
in][]{2008PhRvD..78h3519M}.  While the relation between our best-fit
$\fp{}$ and halo mass is close to that obtained from the
peak-background split the values of $\fpp{}$ can differ significantly.
Imposing the peak-background split value of $\fpp{}$ has only a modest
effect on the shape of the correlation function on the scales of
interest however, and does not change our conclusions in any
qualitative way.  $\fpp{}$ is also not well constrained in our fitting
(see Table \ref{table:bias-parameters} for the uncertainty of $\fpp{}$),
which confirms that $\fpp{}$ does not have a considerable impact on the
correlation function on the scales of interest.

Fig.~\ref{fig:real-space-xi} compares the real-space correlation
function predicted by CLPT with that measured in the simulations.
Note that the consistency between CLPT results and simulations is
almost perfect from $\lsim 10\Mpch$ through the BAO scale ($\sim
110\Mpch$), as also seen in \citet{2012arXiv1209.0780C}.  The
redshift-space correlation function predicted directly from CLPT was
presented in \citet{2012arXiv1209.0780C}.  Here we want to examine the
velocity statistics themselves.

We assume that the values of $\fp{}$ and $\fpp{}$ obtained by fitting
the real-space correlation functions are the right ones for evaluating
the velocity statistics.  Using these values in
Eq.~\eqref{eq:v-kernel-m-2nd} and \eqref{eq:vd-kernel-m-2nd}, we
obtain the scale dependence of the pairwise infall velocity and
velocity dispersion.  The CLPT results (divided by the linear theory
as fiducial values) are compared with simulations in
Figs.~\ref{fig:real-space-v12} and \ref{fig:real-space-sigma12}.  The
CLPT predictions for $v_{12}$ are better than the SPT predictions with
first order bias presented in \citet{2011MNRAS.417.1913R} for all but
the highest mass bin.  The CLPT predictions of the pairwise infall
velocity statistics can be slightly improved by varying $\fpp{}$,
but the prediction of velocity dispersion is quite insensitive to
$\fpp{}$.

When comparing the CLPT result to $\vd{\parallel}(r)$ and
$\vd{\bot}(r)$ we add a constant to the predictions so that they take
the same value as the simulation at $r=130\Mpch$.  The constant
offsets for $\vd{\parallel}(r)$ and $\vd{\bot}(r)$ are almost the
same, with only $\sim 1$ per cent relative difference.  These two
constants are similar to what the authors referred to in
\citet{2011MNRAS.417.1913R}: the CLPT prediction of absolute value of
$\vd{\parallel}(r)$ and $\vd{\bot}(r)$ is not correct, but a constant
shift over the whole range of scales reveals that the CLPT results
have correct trend.  The possible reason for this is that the velocity
dispersion component yielded by gravitational evolution on smallest
scales, which should be separated from the overall scale dependence of
pairwise velocity dispersion, is not able to be predicted by
perturbation theory: this suggests that we should evaluate the
constant shift as a fitting parameter.  Our CLPT predictions have a
similar accuracy to the SPT predictions in \citet{2011MNRAS.417.1913R}
for the second highest mass bin.  However, it is not clear that CLPT
accurately captures the bias-dependence of the deviations from linear
theory for $\vd{\parallel}(r)$ and $\vd{\bot}(r)$.

\subsubsection{Redshift-space distortion for auto-correlation}
\label{sec:redshift-space-stat}

\begin{figure}
  \centering
  \includegraphics[width=80mm, keepaspectratio]{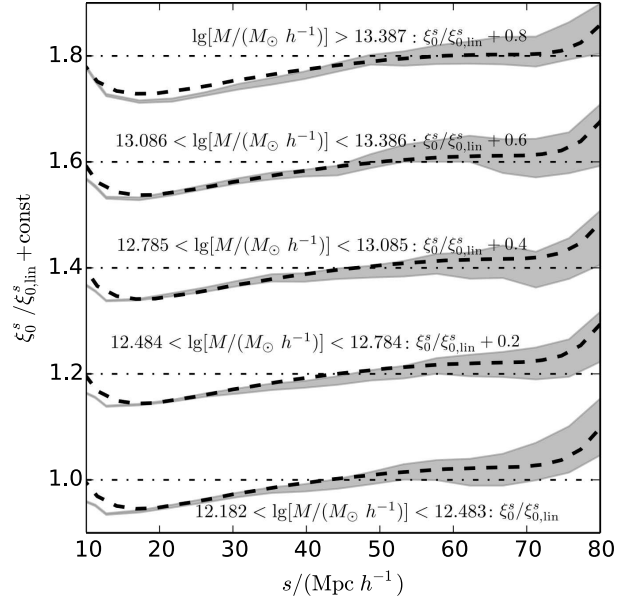}
  \caption{Redshift-space statistics obtained by Gaussian streaming
    model specified in \citet{2011MNRAS.417.1913R}, showing
    $\xi^s_0/\xi^s_{0,\mathrm{lin}}$. The shaded bands present
    simulation values (showing the error range) and our CLPT values
    are presented by heavy dashed curve. Each mass bin is elevated by
    a different constant.}
  \label{fig:redshift-monopole}
\end{figure}

\begin{figure}
  \centering
  \includegraphics[width=80mm, keepaspectratio]{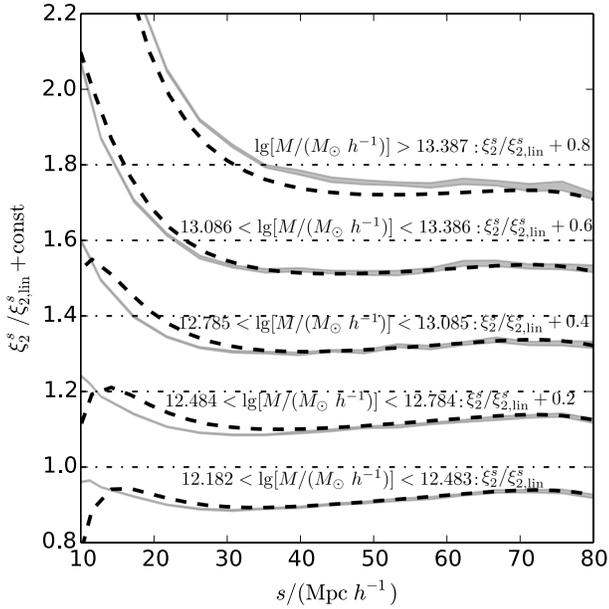}
  \caption{Redshift-space results of $\xi^s_2/\xi^s_{2,\mathrm{lin}}$.
    Labels and curve indications are identical to
    Fig.~\ref{fig:redshift-monopole}.}
  \label{fig:redshift-quadrupole}
\end{figure}

The redshift-space correlation functions depend on the angle between
separation vectors and LOS. This ``direction dependency'' can be
expanded into series with respect to Legendre polynomials
(Eq.~\ref{eq:legendre-expansion}), or, equivalently:
\begin{equation}
  \label{eq:multipole-expansion}
  \xi^s_l(s) = \dfrac{2l+1}{2}
  \int_{-1}^1\d\mu_s\ L_l(\mu_s) \xi^s_l(s,\mu)\ ,
\end{equation}
where $s = (r_p^2 + r_\parallel^2)^{1/2}$ is the redshift-space
distance, and $\mu_s = r_\parallel / s$ is the cosine of the angle
between separation vector and LOS. Generally we are most interested in
the lowest non-zero moments, i.e. $l = 0,\ 2$. In
Figs.~\ref{fig:redshift-monopole} and \ref{fig:redshift-quadrupole},
we present the lowest two non-zero multipole moments divided by linear
theory results \citep[see] [for the linear theory expessions]
{1995ApJ...448..494F, 2011MNRAS.417.1913R}.

We show the results of the multipole expansion in
Figs.~\ref{fig:redshift-monopole} and \ref{fig:redshift-quadrupole}.
The accuracy of the prediction of CLPT with the Gaussian streaming
model is at the several per cent level on scales larger than $\gsim
20\Mpch$, and no worse than $\sim 10$ per cent even down to $\sim
10\Mpch$.  The agreement remains equally good at BAO scales, but we
only show $10\Mpch < s < 70\Mpch$ for clearer presentation at smaller
scales.  The theory breaks down at $r\lsim 10\Mpch$ where the
correlation function amplitude is approaching $O(1)$.  For the
quadrupole moment ($\ell=2$), we observe that the model has reasonable
overlap with the simulations throughout the scales of general interest
($10\Mpch \lsim s \lsim 120\Mpch$).

\begin{figure}
  \centering
  \includegraphics[width=80mm, keepaspectratio]
  {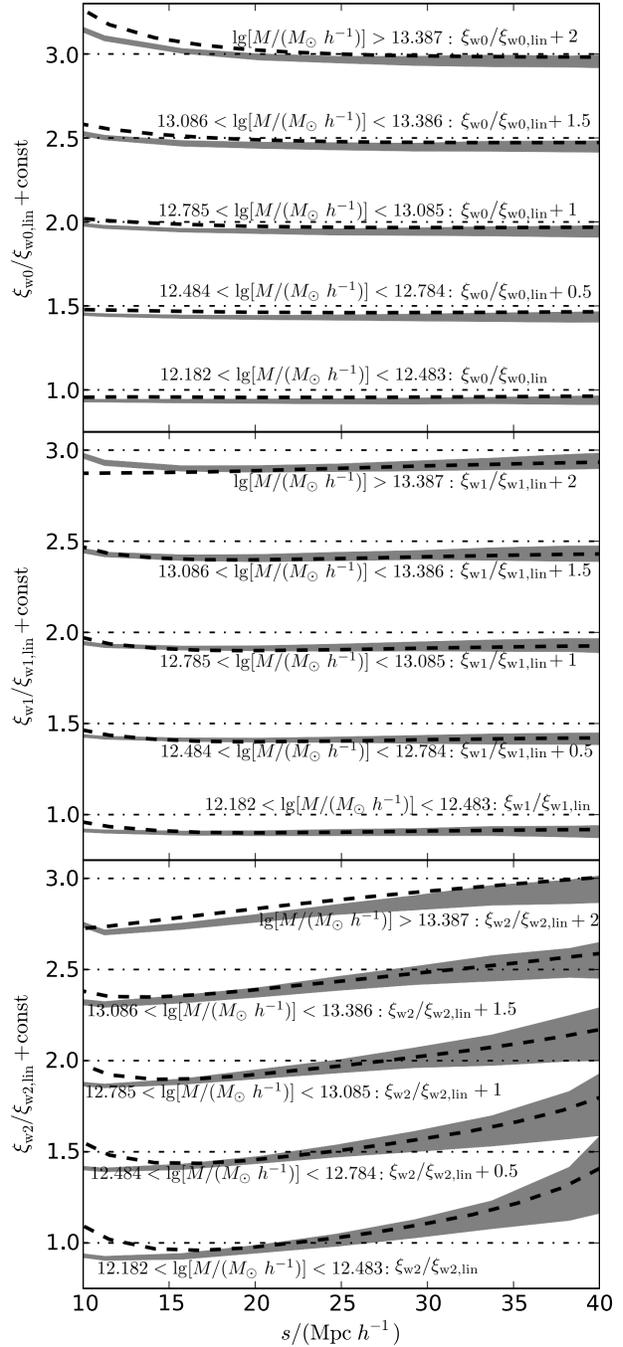}
  \caption{Wedge statistics showing $\xi_\mathrm{w0}$,
    $\xi_\mathrm{w1}$ and $\xi_\mathrm{w2}$ (see
    Eq.~\ref{eq:wedge-definition}) in redshift-space. The scheme of
    presentation is similar to Fig.~\ref{fig:redshift-monopole}.}
  \label{fig:wedge}
\end{figure}

To further isolate the regions where the theory and N-body simulations
are in good agreement we additionally examine the ``wedge'' statistics
\citep[e.g.][]{2012MNRAS.419.3223K}, defined by
\begin{equation}
  \label{eq:wedge-definition}
  \xi^s_\mathrm{wedge}(s,\mu_\mathrm{min},\mu_\mathrm{max}) =
  \dfrac{1}{\Delta\mu} \int_{\mu_\mathrm{min}}^{\mu_\mathrm{max}}
  \xi^s(s,\mu) \d \mu\ ,
\end{equation}
where $\Delta\mu=\mu_\mathrm{max}-\mu_\mathrm{min}$.  In this paper we
use three such ``wedges'', which are denoted by
$\xi_\mathrm{w0}=\xi^s_\mathrm{wedge}(s,0,1/3)$,
$\xi_\mathrm{w1}=\xi^s_\mathrm{wedge}(s,1/3,2/3)$ and
$\xi_\mathrm{w2}=\xi^s_\mathrm{wedge}(s,2/3,1)$.  The predictions for
the $\xi_{\mathrm{w}i}$ are compared to N-body simulations in
Fig.~\ref{fig:wedge}.  Note that the fractional deviations from linear
theory are largest on small scales and when $\mu\simeq 1$.  In
addition the inaccuracy of our theoretical prediction for the
quadrupolar moment on about $\sim 10\Mpch$ can be attributed to the
disagreement near $\mu\simeq 1$ (please note that $\xi^s_2$ is
negative around $r\sim 10\Mpch$ but $\xi_\mathrm{w2}$ is positive
there).  On scales above $20\Mpch$ our model works well, the
difference between the model and N-body results is less than 5 per
cent for all three wedges.

\begin{figure}
  \centering
  \includegraphics[width=80mm, keepaspectratio]{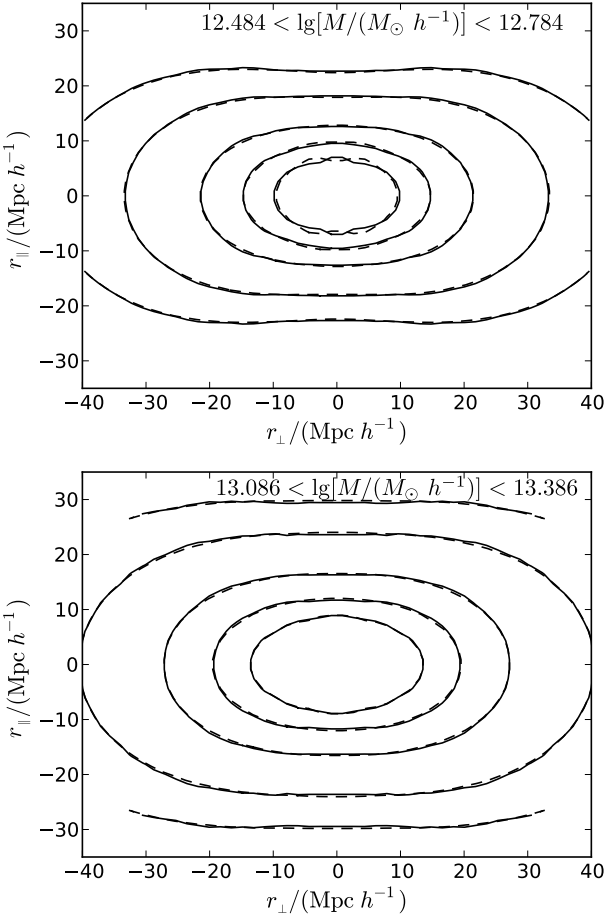}
  \caption{Contour plots that compare theoretical predictions (dashed)
    and simulation results (solid). The contour lines for $\xi^s=0.8,\
    0.4,\ 0.2,\ 0.08\ ,$ and $0.04$ are presented in the figure. We
    note that the dashed and solid contours are nicely overlapped in
    most areas, except in the regions that $s \lsim 10\ \Mpch$ and
    $\mu \lsim 1$. }
  \label{fig:2d-contour}
\end{figure}

In order to provide another view of the disagreement between the model
and simulations, we show in Fig.~\ref{fig:2d-contour} contours of
$\xi$ predicted by the analytic model (dashed contours) and N-body
simulations (solid contours) for two bins in halo mass.  We can
clearly observe that, for the halos in the lower mass bin, $\xi^s$ is
less precisely predicted around $\mu\simeq 1$. On larger scales
($s\gsim 20\Mpch$) simulation results are accurately predicted for
both mass bins.

\subsection{Cross correlation between halos and dark matter particles}
\label{sec:cross-corr-halo-dm}

As shown in subsection \ref{sec:cross-corr-theory}, CLPT theory is
also capable of making predictions for cross-correlations.  Here we
compare statistics predicted by CLPT and the Gaussian streaming model
with those given by simulations, for cross-correlations of halos with
dark matter particles in the simulations.
In this subsection we adopt the same bias parameters as in Table
\ref{table:bias-parameters}.
We can obtain a better match to the cross-correlation infall velocities
by adjusting $\fp{}$ and $\fpp{}$, however those values do not provide a
good match to the real-space cross-correlation function suggesting either
that our bias model is too simple or the improved agreement reflects a
breakdown of perturbation theory.

\begin{figure}
  \centering
  \includegraphics[width=80mm, keepaspectratio]
  {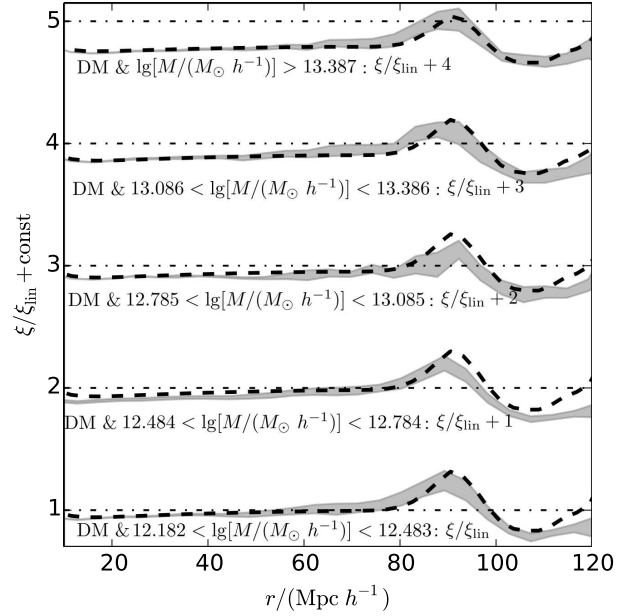}
  \caption{Real-space cross-correlation function between dark matter
    and halos (in five different mass bins; each mass bin is elevated
    by a specific constant). The scheme of presentation is similar to
    Fig.~\ref{fig:real-space-xi}.}
  \label{fig:real-space-xi-cross}
\end{figure}

\begin{figure} \centering
  \includegraphics[width=80mm, keepaspectratio]
  {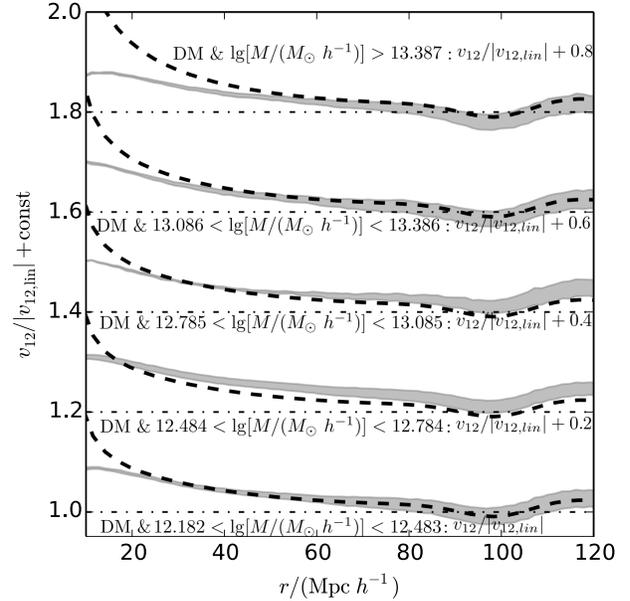}
  \caption{Pairwise infall velocity as cross-correlations. Our CLPT
    results are shown by a heavy solid curve, compared with
    simulations presented by shaded bands showing the error range.
    The scheme of presentation is similar to
    Fig.~\ref{fig:real-space-v12}.}
  \label{fig:real-space-v12-cross}
\end{figure}

\begin{figure}
  \centering
  \includegraphics[width=80mm, keepaspectratio]
  {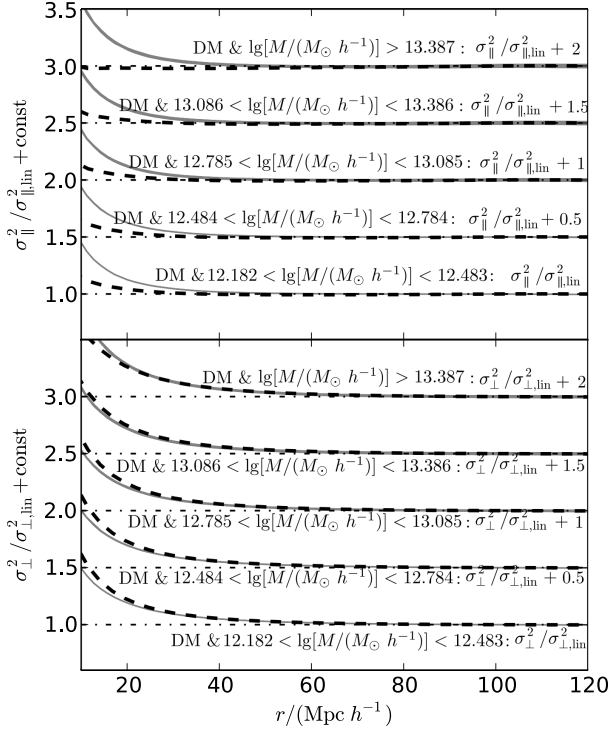}
  \caption{Pairwise velocity dispersion as cross-correlations.
    Similar to Fig.~\ref{fig:real-space-sigma12}, we also show the
    values of $\vd{\parallel} / \vd{\parallel,\mathrm{lin}}$ (upper
    panel) and $\vd{\bot} / \vd{\bot,\mathrm{lin}}$ (lower panel),
    with shaded bands (simulation results) and dashed curves
    (theoretical predictions). Different mass bins are elevated by
    different constants. }
  \label{fig:real-space-sigma12-cross}
\end{figure}

Real-space statistics are presented by
Figs.~\ref{fig:real-space-xi-cross} through
\ref{fig:real-space-sigma12-cross}. Predictions of CLPT for pairwise
infall velocity (Fig.~\ref{fig:real-space-v12-cross}) in such a
cross-correlation case is not as good as the prediction for
auto-correlation (Fig.~\ref{fig:real-space-v12}), but still
satisfactory; the discrepancy is $\lsim 5$ per cent throughout the
scales of interest.  The real-space correlation function, $\xi(r)$,
and the velocity dispersion, $\vd{12}(r)$, on the other hand, are
still accurately predicted by the theory.


\begin{figure}
  \centering
  \includegraphics[width=80mm, keepaspectratio]
  {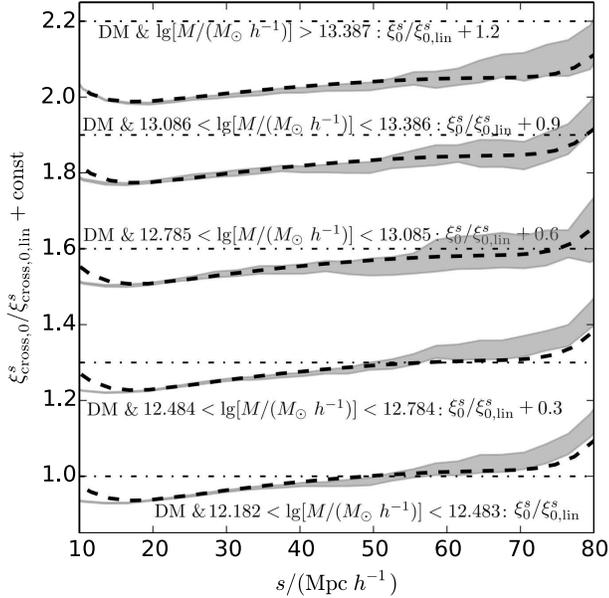}
  \caption{Monopole moment of redshift-space cross-correlation
    function between halos and dark matter. Results are divided by
    fiducial linear theory results (i.e. $\xi^s_0/\xi^s_{0,lin}$ is
    presented). The scheme of presentation is similar to
    Fig.~\ref{fig:redshift-monopole}.}
  \label{fig:redshift-monopole-cross}
\end{figure}

\begin{figure}
  \centering
  \includegraphics[width=80mm, keepaspectratio]
  {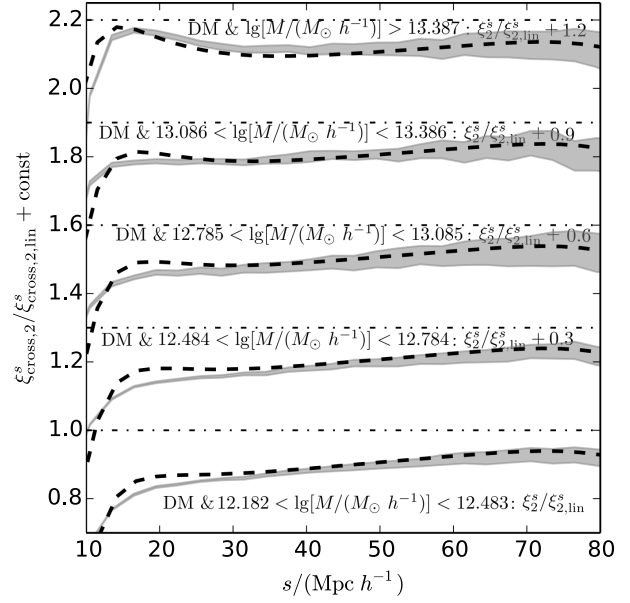}
  \caption{Quadrupole moment of redshift-space cross-correlation
    function between halos and dark matter (linear theory results as
    fiducial value) $\xi^s_2/\xi^s_{2,lin}$. The scheme of
    presentation is similar to
    Fig.~\ref{fig:redshift-monopole-cross}.}
  \label{fig:redshift-quadrupole-cross}
\end{figure}

Inserting the real-space statistics shown by
Figs.~\ref{fig:real-space-xi-cross} through
\ref{fig:real-space-sigma12-cross} into Eq.~\eqref{eq:streaming-xi-s},
we get the redshift-space correlation function, which is also expanded
with respect to Legendre polynomials as in
Eq.~\eqref{eq:multipole-expansion}. Similar to Section
\eqref{sec:redshift-space-stat}, this section also presents monopole
and quadrupole moments in Figs.~\ref{fig:redshift-monopole-cross} and
\ref{fig:redshift-quadrupole-cross}. While we focus on $10\Mpch < s <
70\Mpch$, the agreement remains good on BAO scales.  Although the
predictions for the velocity statistics are not as good as in the
auto-correlation case, the behavior of the multipole moments is still
well sketched by CLPT and the Gaussian streaming model (to $\sim 10$
per cent, even on scales of $\sim 10\Mpch$).  The manner in which
$\xi^s_0/\xi^s_\mathrm{0,lin}$ and $\xi^s_2/\xi^s_\mathrm{2,lin}$ vary
with $s$ is still correct to quite small $s$.

\begin{figure}
  \centering
  \includegraphics[width=80mm, keepaspectratio]
  {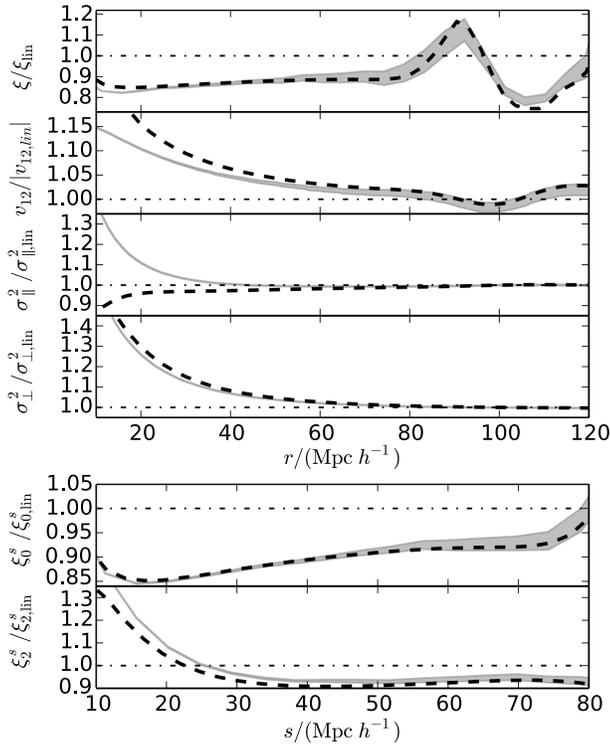}
  \caption{Statistic functions as cross-correlations betweeen halos in
    two different mass bins: $12.182 < \lg[M/(M_\odot h^{-1})] <
    12.483$ and $\lg[M/(M_\odot h^{-1})] > 13.387$. The curvs, lines
    and shaded bands in the panels have similar indications to
    Figs.~\ref{fig:real-space-xi} through
    \ref{fig:redshift-quadrupole}, which compare our theoretical
    predictions with simulations. From top to bottom: real-space
    correlation function (first panel); pairwise infall velocity
    (second panel); velocity dispersion parallel (third panel) and
    perpandicular (fourth panel) to separation vector; monopole (fifth
    panel) and quadrupole (sixth panel) moment of redshift-space
    correlation function.}
  \label{fig:halo-halo-cross}
\end{figure}

Similar procedures also produce predictions for halo-halo cross
correlations. In Fig.~\ref{fig:halo-halo-cross} we compare the
statistics as cross-correlations betweeen halos in two different
mass bins: $12.182 < \lg[M/(M_\odot h^{-1})] < 12.483$ and
$\lg[M/(M_\odot h^{-1})] > 13.387$.  There are no adjustable parameters
in this comparison, because the values of $\fp{}$ and $\fpp{}$ are fixed
by the auto-correlations.
Good agreement between our theoretical model and the simulations is
still observed, even down to the scale of $s\gsim 10\Mpch$.

\section{Discussion and summary}
\label{sec:summary}

By introducing an auxiliary term $\vec{J}$ in the generating function,
we generalize the CLPT scheme elaborated in
\citet{2012arXiv1209.0780C} to estimate the pairwise infall velocity
and velocity dispersion as functions of pair separation.  This allows
a self-consistent calculation of these statistics for biased tracers,
including scale-dependent or higher-order bias terms.  Indeed we find
that CLPT gives better estimates for the magnitude of the pairwise
infall velocity, $v_{12}(r)$, than to quasi-linear theory with a
``linear'' bias \citep{2011MNRAS.417.1913R} for a wide range of halo
masses.

The $\xi(r)$, $v_{12}(r)$ and $\vd{12}(r)$ predicted by CLPT can be
used as inputs to Gaussian streaming model
(Eq.~\ref{eq:streaming-xi-s}) to obtain predictions for the
redshift-space correlation function of halos.  For the monopole and
quadrupole moments of the correlation function the agreement between
theory and N-body simulations is at the few per cent level down to
$\sim 15\Mpch$, and $\sim 10$ per cent at $\sim 10\Mpch$.  We infer
that the Gaussian streaming model of redshift-space distortion is not
sensitively affected by $\vd{12}(r)$, but the small scale statistics
are enhanced by better estimations of $v_{12}(r)$, compared with
semi-linear results in \citet{2011MNRAS.417.1913R}. We attribute the
enhanced results to our inclusion of higher order (one-loop) terms and
the resummation scheme employed in CLPT.

It is worth noting that the argeement between the CLPT-Gaussian
streaming quadrupolar moment and N-body simulation
(e.g.~Fig.~\ref{fig:redshift-quadrupole}) is consideribaly better than
the ``original'' SPT scheme in \citet{2011MNRAS.417.1913R}.  From
Figs.~\ref{fig:wedge} and \ref{fig:2d-contour} we observe that the
theoretical predictions are not sufficiently accurate only around the
region where $s\lsim 20\Mpch$ and $\mu\simeq 1$ (or $r_p\simeq 0$) for
the lower mass bins.  It was shown in \citet{2011MNRAS.417.1913R}
(e.g.~their figure 6) that the Gaussian streaming model predicted the
redshift-space correlation function well when accurate ``inputs''
(i.e. $v_{12}(r)$ and $\vd{12}(r)$) were used.  Prediction of these
inputs using CLPT seems a reliable way of computing redshift-space
statistics for tracers with a local Lagrangian bias.

We also extended the CLPT-Gaussian streaming model to
cross-correlations between differently biased tracers.  As an example,
we modelled the monopole and quadrupole moments of the redshift-space
cross-correlation function between halos and dark matter, and
between halos in different mass bins.  The agreement with
N-body simulations for $v_{12}(r)$ was not as good as in the
auto-correlation case, as expected, but the distortions were still
accurately revealed in monopole and quadrupole moments.  This is not
unexpected: it was already noted by \citet{2011MNRAS.417.1913R} that
the Gaussian approximation for the velocity PDF worked much better for
halos in simulations than for the dark matter particles themselves.

\bibliographystyle{mn2e}
\bibliography{ms}
\end{document}